\documentclass[twocolumn,superscriptaddress,showpacs,prl,aps,amsmath,amssymb,nofootinbib]{revtex4-1} 
\usepackage{natbib}
\usepackage{graphicx,color}
\usepackage{amsmath,amssymb}
\usepackage{verbatim}
\usepackage{float}
\usepackage{wasysym}
\usepackage{amssymb,graphicx}
\usepackage{epsfig}
\usepackage{psfrag}
\usepackage{dsfont}
\usepackage{amsfonts}
\usepackage{mathrsfs}
\usepackage{multirow}
\usepackage{times}
\usepackage{bm}
\usepackage{hyperref}
\hypersetup{
  colorlinks=true,        
  linkcolor=blue,         
  citecolor=cyan,         
}
\usepackage{blindtext, rotating}
\usepackage{pifont}
\usepackage[normalem]{ulem}   
\usepackage{grffile}
\graphicspath{{./}}

\newcommand{\beq}{\begin{equation}} 
\newcommand{\eeq}{\end{equation}} 
\newcommand{\beqn}{\begin{eqnarray}} 
\newcommand{\eeqn}{\end{eqnarray}}

\newcommand{\zD}{{\raise1.0ex\hbox{${}^{\ \circ}$}}\!\!\!\!\!D}
\newcommand{\alone}{{\raise0.5ex\hbox{${}^{\ 1}$}}\!\!\!\!\alpha}

\newcommand{\nalam}{\mathrel{\raise0.9ex\hbox{$^\lambda$}\mkern-14mu
\lower0.0ex\hbox{$\nabla$}}}

\newcommand{\zeroD}{{\raise1.0ex\hbox{${}^{\ \circ}$}}\!\!\!\!\!D}

\newcommand{\zLap}{{\raise1.0ex\hbox{${}^{\ \circ}$}}\!\!\!\!\Delta}
\newcommand{\zna}{{\raise1.0ex\hbox{${}^{\ \circ}$}}\!\!\!\!\!\nabla}
\newcommand{\zS}{{\raise1.0ex\hbox{${}^{\ \circ}$}}\!\!\!\!\!S}

\newcommand{\GA}{\alpha}

\newcommand{\GG}{\gamma}

\newcommand{\GE}{\epsilon}

\newcommand{\GR}{\rho}
\newcommand{\GS}{\sigma}

\newcommand{\GU}{\theta}

\newcommand{\Ms}{M_\odot}
\newcommand{\cmark}{\ding{51}}%
\newcommand{\xmark}{\ding{55}}%

\newcommand{\be}{\begin{equation}}
\newcommand{\ee}{\end{equation}}

\def\QEQ{{%
    \setbox0\hbox{$I$}%
    \rlap{\hbox to \wd0{\hss--\hss}}\box0
}}

\begin{document}
\title{Magnetic Ergostars, Jet Formation and Gamma-Ray Bursts: Ergoregions versus Horizons}
\author{Milton Ruiz}
\affiliation{Department of Physics, University of Illinois at
  Urbana-Champaign, Urbana, Illinois 61801}
\author{Antonios Tsokaros}
\affiliation{Department of Physics, University of Illinois at
  Urbana-Champaign, Urbana, Illinois 61801}
\author{Stuart L. Shapiro}
\affiliation{Department of Physics, University of Illinois at
  Urbana-Champaign, Urbana, Illinois 61801}
\affiliation{Department of Astronomy \& NCSA, University of
  Illinois at Urbana-Champaign, Urbana, Illinois 61801}
\author{Kyle C. Nelli}
\affiliation{Department of Physics, University of Illinois at
  Urbana-Champaign, Urbana, Illinois 61801}
\author{Sam Qunell}
\affiliation{Department of Physics, University of Illinois at
  Urbana-Champaign, Urbana, Illinois 61801}
\date{\today}

\begin{abstract}
  We perform the first fully general relativistic, magnetohydrodynamic simulations of dynamically stable
  hypermassive neutron stars with and without ergoregions to assess the impact of ergoregions on launching
  magnetically--driven outflows. The hypermassive neutron stars are modeled by a compressible and causal
  equation of state and are initially endowed with a dipolar magnetic field extending from the stellar
  interior into its exterior. We find that, after a few Alfv\'en times, magnetic field lines in the ergostar
  (star that contains ergoregions) and the normal star, have been tightly wound in both cases into a helical
  funnel within which matter begins to flow outward.  The maximum Lorentz factor in the outflow is $\Gamma_L
  \sim 2.5$, while the force-free parameter holds at $B^2/8\pi\rho_0\lesssim 10$. These values are
  incompatible with highly relativistic, magnetically-driven outflows (jets) and short $\gamma$-ray bursts. 
  We compare these results with those of a spinning black hole surrounded by a magnetized, massless 
  accretion disk that launches a bona fide jet. Our simulations suggest that the Blandford-Znajek
  mechanism  for launching relativistic jets only operates  when a black hole is present, though the Poynting
  luminosity in all cases is comparable. Therefore, one cannot distinguish a magnetized, accreting black hole
  from a magnetized hypermassive neutron star in the so-called mass-gap based solely on the value of the
  observed Poynting luminosity. These results complement our previous studies of supramassive remnants and
  suggest that it would be challenging for either normal neutron stars or ergostars in a hypermassive state
  to be the progenitors of short $\gamma$-ray bursts.     
\end{abstract}
\maketitle

\textit{Introduction.}\textemdash
Event GW170817~\cite{TheLIGOScientific:2017qsa} marked not only the first direct detection
of a binary neutron star (BNS) merger via gravitational waves but also the
simultaneous detection of the short $\gamma$-ray burst (sGRB) GRB170817A
\cite{2017GCN.21520....1V,2017GCN.21517....1K}, and kilonova AT 2017gfo, with its afterglow
radiation in the radio, optical/IR, and X-ray bands. These detections constituted a golden moment in the era of
multimessenger astronomy~\cite{GBM:2017lvd,Monitor:2017mdv, Abbott:2017wuw}. To investigate
the different scenarios for jet formation triggering the such electromagnetic (EM) events, fully general relativistic,
magnetohydrodynamic (GRMHD) simulations are needed (for recently reviews see~e.g.
\cite{Paschalidis:2016agf,Baiotti:2016qnr,Ciolfi:2020cpf}). 

The most common scenario for launching a magnetically-driven jet
is a black hole-disk (BH-disk) remnant. BNS mergers that lead to hypermassive remnants (HMNSs) inevitably
collapse to BHs immersed in gaseous disks. These remnants are robust
engines for jet launching~\cite{prs15,Ruiz:2018wah,Ruiz:2016rai,Ruiz:2017inq,Ruiz:2019ezy}.
Their lifetimes are $\Delta t\simeq 150\ {\rm ms}$ and outgoing Poynting luminosities
$\sim 10^{52\pm 1}\ \rm erg/s$, both 
consistent with typical sGRBs~\cite{Bhat:2016odd,Lien:2016zny,Svinkin:2016fho,Ajello:2019zki}, 
as well as with the Blandford-Znajek (BZ) mechanism for launching jets and their associated
Poynting luminosities \cite{BZeffect77,Thorne86}. The key requirement for the emergence of a jet is
the existence of a large-scale poloidal magnetic field along the direction of the total orbital
angular momentum of the BH-disk remnant~\cite{Ruiz:2020via}. Such a field arises when the NS is initially endowed
with a dipolar magnetic field confined or not to the interior of the NSs.
During the BNS merger and the HMNS phase, magnetic 
instabilities drive the magnetic energy to saturation levels~\cite{Kiuchi:2014hja}.
Following the HMNS collapse to a BH, such magnetic fields launch a mildly relativistic,
magnetically-driven outflow with a Lorentz factor $\Gamma_L\gtrsim 1.2$ confined inside a
tightly-wound-magnetic funnel. This becomes an ``incipient jet''
once regions above the BH poles approach force-free values ($B^2/8\,\pi \rho_0\gg 1$).
Here $B$ and $\rho_0$ are the strength of the magnetic field and the rest-mass
density, respectively.  For axisymmetric, Poynting-dominated jets, the maximum
$\Gamma_L$ ultimately attained in the funnel is approximately $B^2/8\,\pi \rho_0$
\citep{B2_over_2RHO_yields_target_Lorentz_factor}. Therefore, incipient jets
will become highly relativistic, as required by sGRB models~\cite{Zou2009}.

The possibility of jet launching from a stable NS remnant has recently been investigated
\cite{Ruiz:2017due,Ciolfi2019,Ciolfi:2020hgg}. 
In particular, Ref. \cite{Ruiz:2017due} presented $200\ \rm ms$-numerical simulations of a stable
(supramassive \cite{1992ApJ...398..203C}) NS remnant initially threaded by a dipolar magnetic 
field that extended from the stellar interior into its exterior. Such a stable NS remnant 
showed no evidence of jet formation, since the outflow confined in the funnel had  $\Gamma_L\lesssim 1.03$ and
 $B^2/8\,\pi \rho_0\lesssim 1$, thereby lacking a 
force-free magnetosphere needed for the BZ--like mechanism to power
a collimated jet \cite{Shapiro:2017cny}. These results suggest that a supramassive NS remnant, which may arise and
live arbitrarily long following the merger of a BNS, probably cannot be the progenitor of a sGRB.
These results have been confirmed in Refs. \cite{Ciolfi2019,Ciolfi:2020hgg}, which
reported the emergence of a $\Gamma_L\lesssim 1.05$-outflow after $\gtrsim 212\ \rm ms$
following the merger of a magnetized low-mass BNS. 
However, it has been suggested that neutrino effects may help reduce the baryon-load
in the region above the poles of the NS, which may drive up the force-free parameter 
in the funnel~\cite{Mosta:2020hlh} and lead to jet formation.

Several  questions need to be addressed
regarding the central engine that launches an incipient jet, as well as the nature
of the jet itself. First, the membrane paradigm implies
that a spinning BH immersed in a disk is a sufficient condition for jet formation~\cite{Thorne82,Thorne86},
but {\it is it also a necessary condition?} If yes, then a stable NS remnant (like a supramassive
NS) cannot be the generator of such jets.
If no, then {\it is a NS jet qualitatively the same as the one launched from BH-disks?}
In particular, {\it can one still describe it as a BZ-like jet?} If not, {\it what are the
  main differences from a BZ-like jet?} It has been argued that contrary to the
membrane paradigm, the horizon is not the ``driving force'' behind the BZ 
mechanism but rather it is the ergoregion \cite{Komissarov:2002dj,Komissarov:2004ms,2005MNRAS.359..801K,
  Ruiz:2012te}. Thus it may be possible
that \textit{normal} NSs cannot launch a BZ jet, but \textit{ergostars} (NSs that contain ergoregions) 
can. If that is the case, {\it can one distinguish between BH-disk and ergostar jets?}

\begin{figure*}
\begin{center}
\begin{turn}{90}  
\bf NS2: Normal star\hspace{0.5cm} $\bm{t/M=630}$
\end{turn}
\includegraphics[width=0.99\columnwidth]{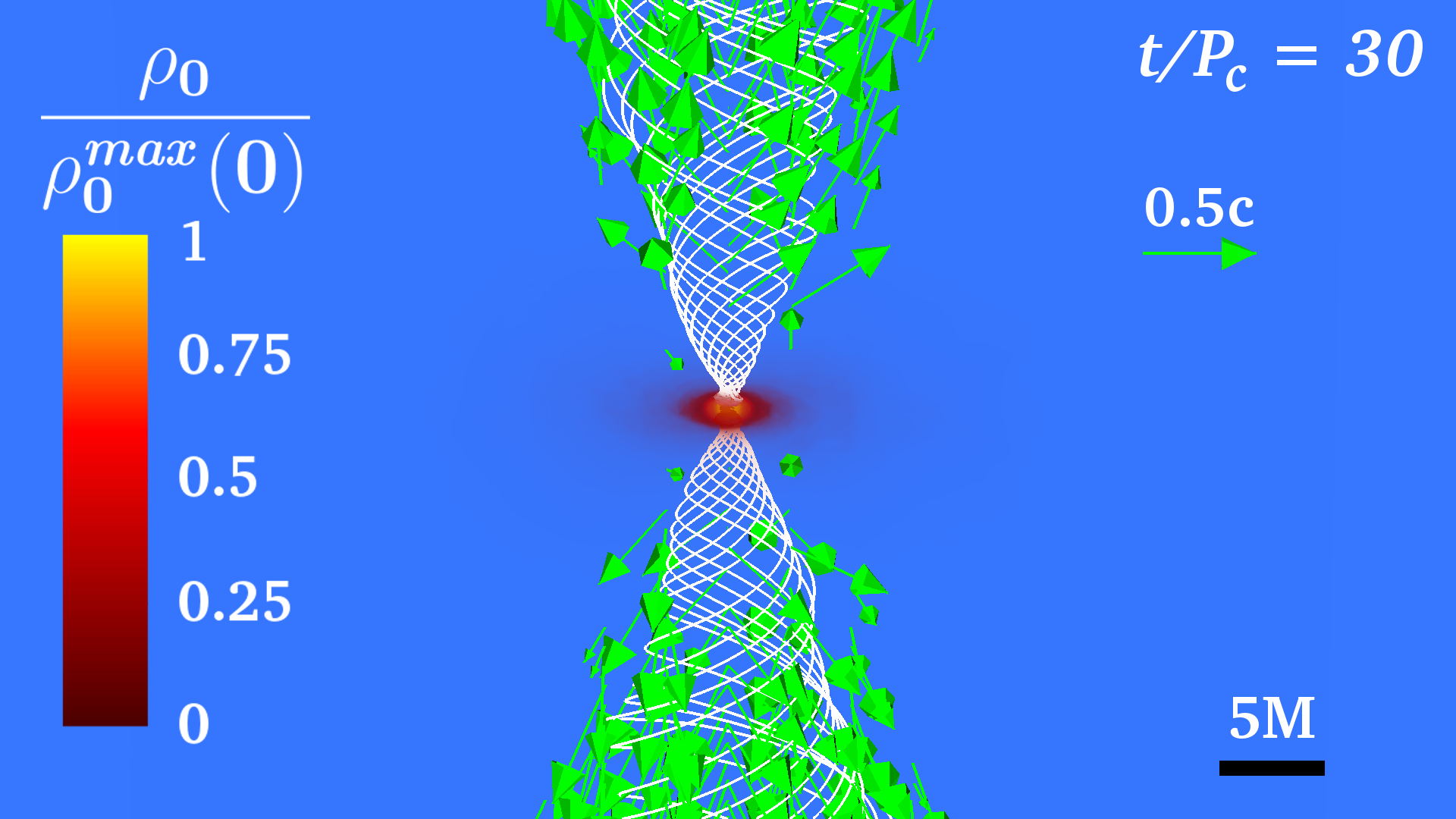}
\includegraphics[width=0.99\columnwidth]{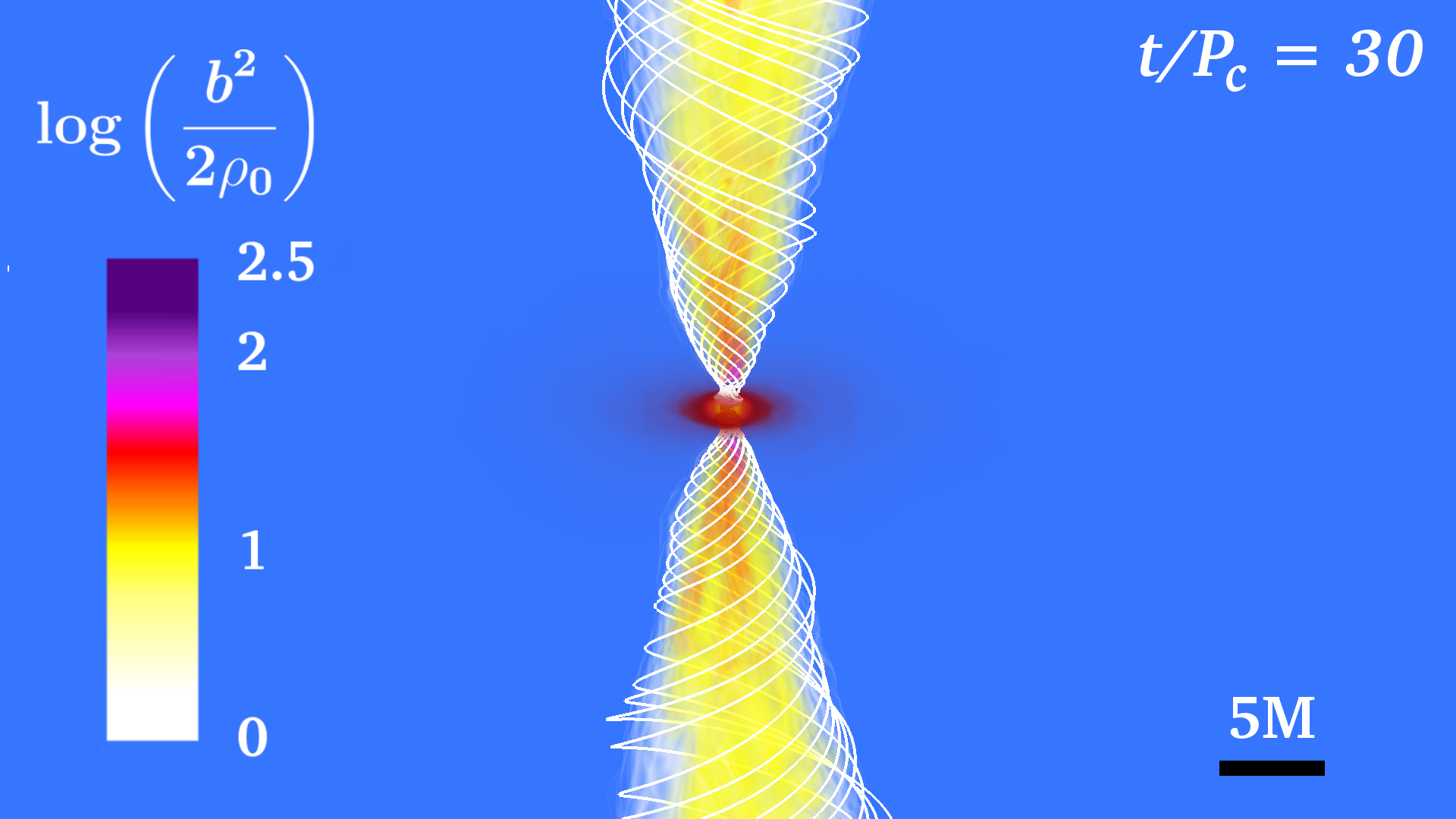}
\vspace{0.1cm}

\begin{turn}{90}  
\bf ES2: Ergostar\hspace{0.5cm} $\bm{t/M=700}$
\end{turn}
\includegraphics[width=0.99\columnwidth]{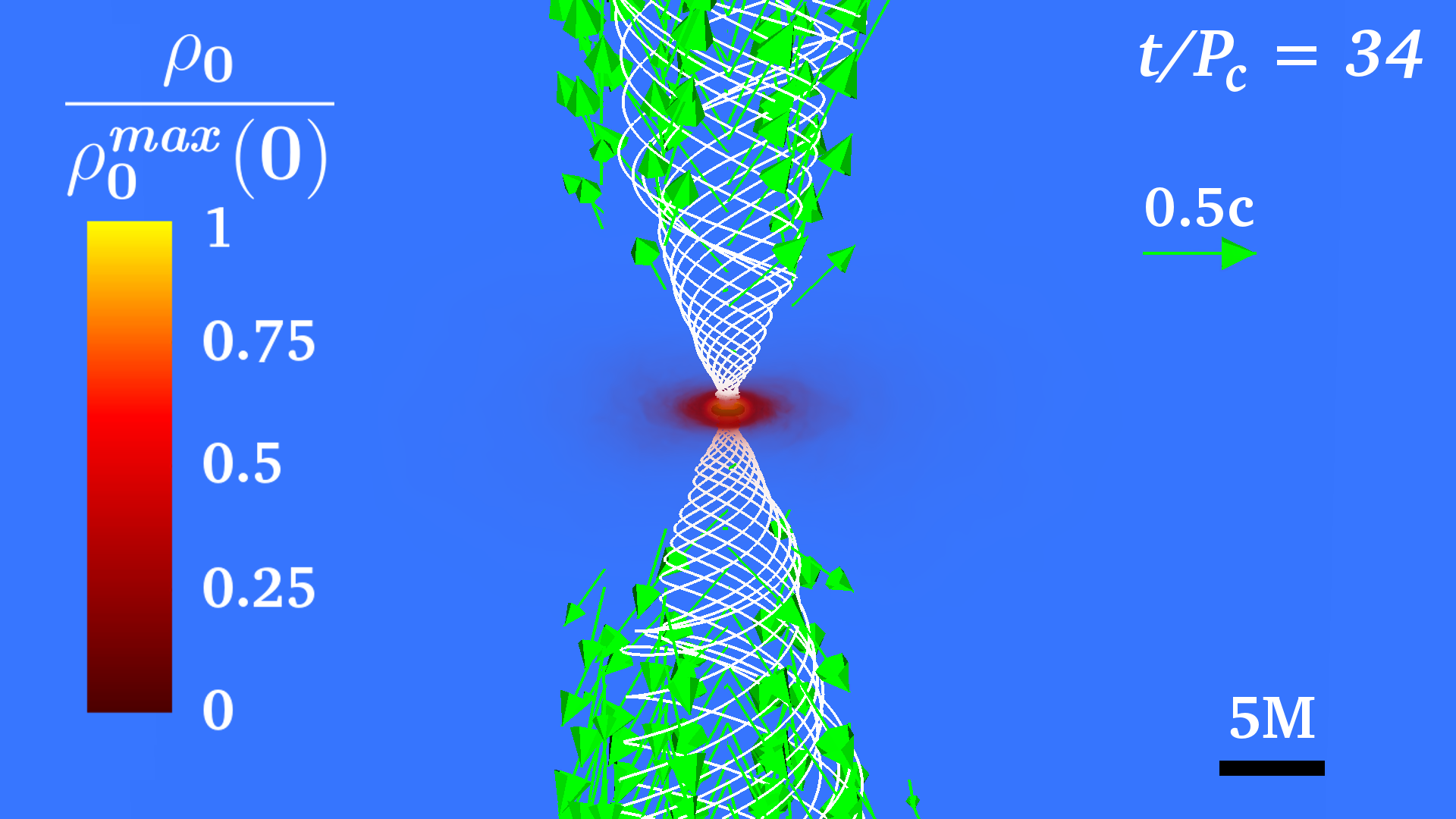}
\includegraphics[width=0.99\columnwidth]{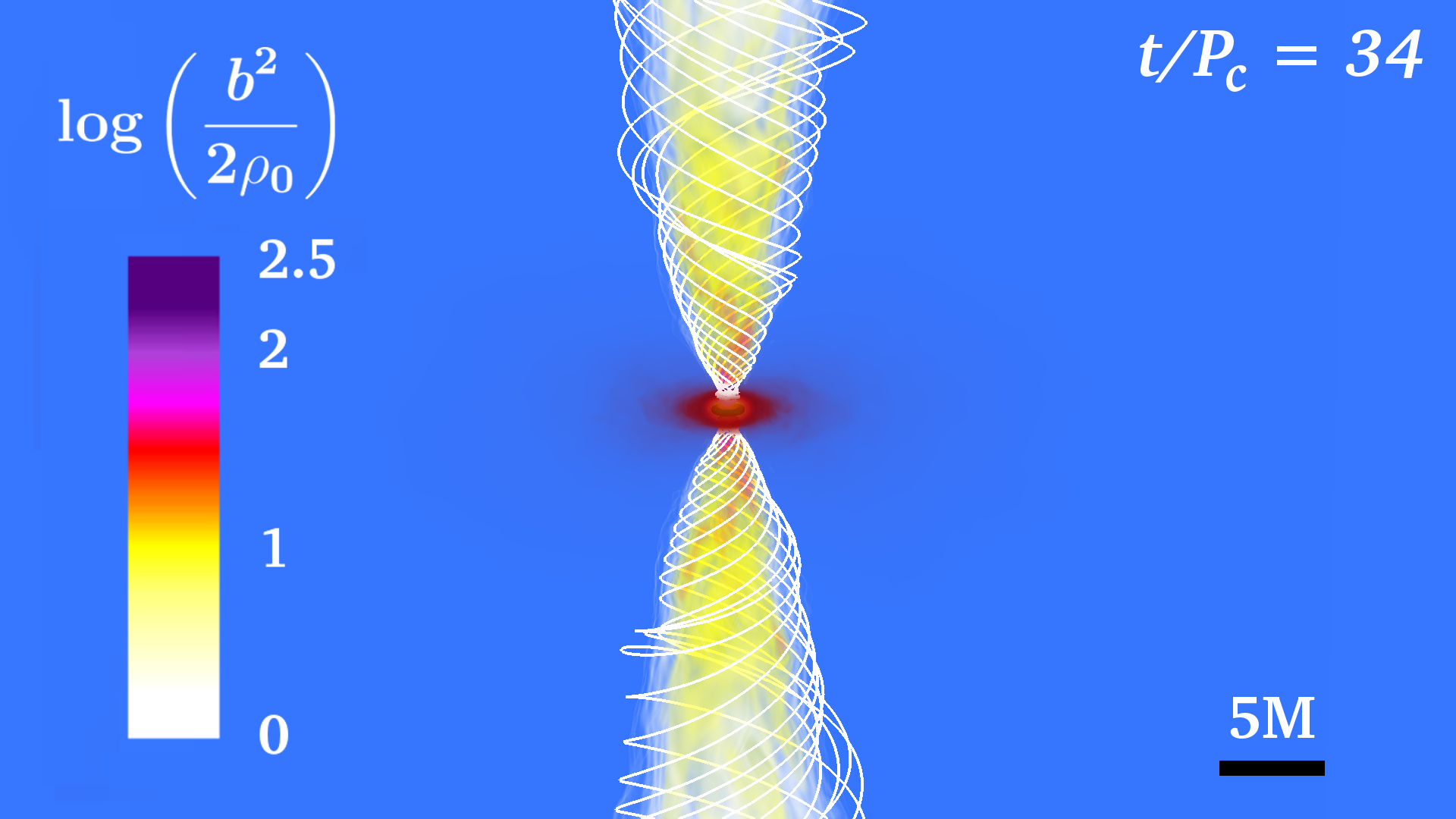}
\vspace{0.1cm}

\begin{turn}{90}  
\bf Black hole-disk \hspace{0.5cm} $\bm{t/M=220}$
\end{turn}
\includegraphics[width=0.99\columnwidth]{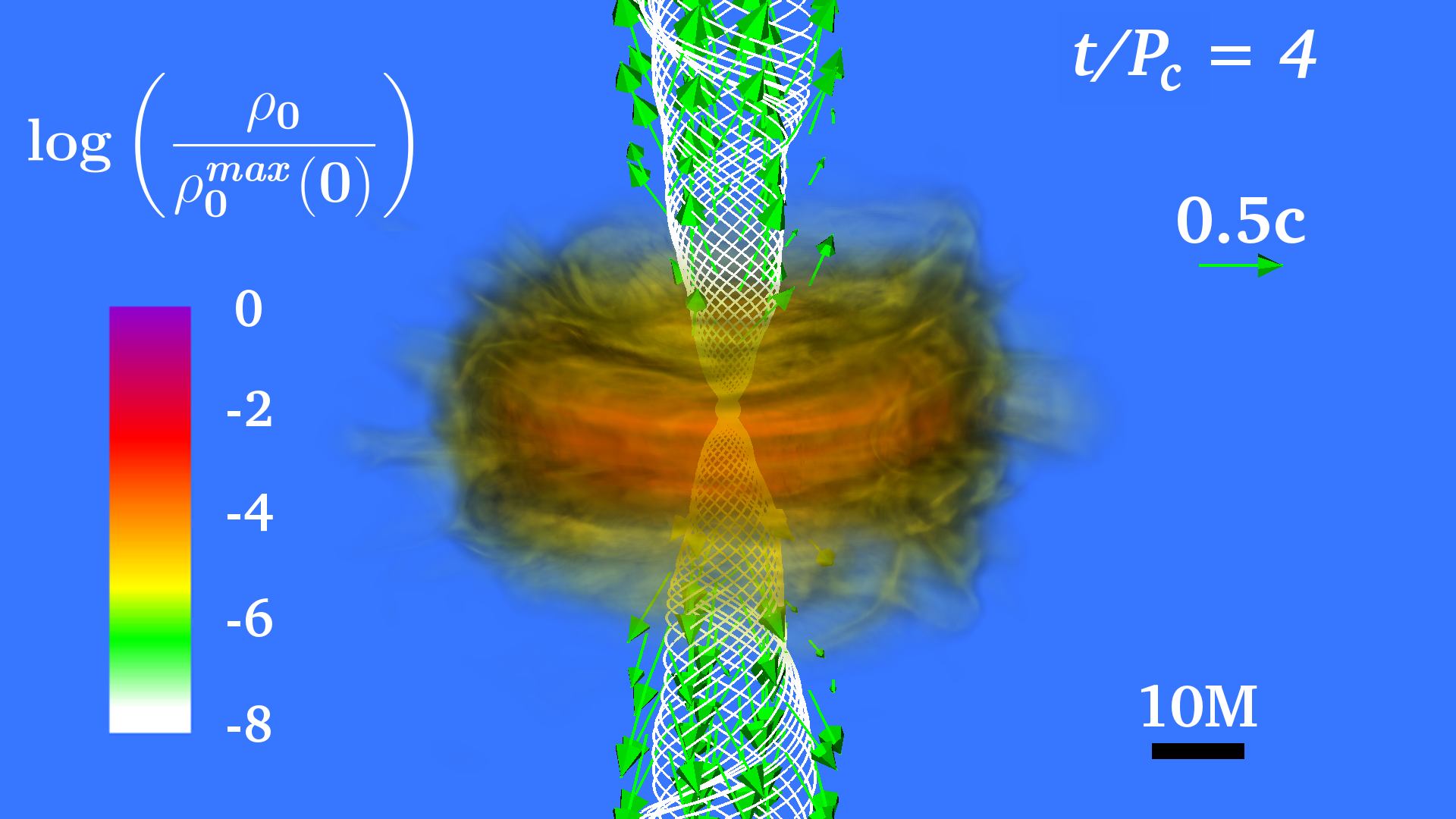}
\includegraphics[width=0.99\columnwidth]{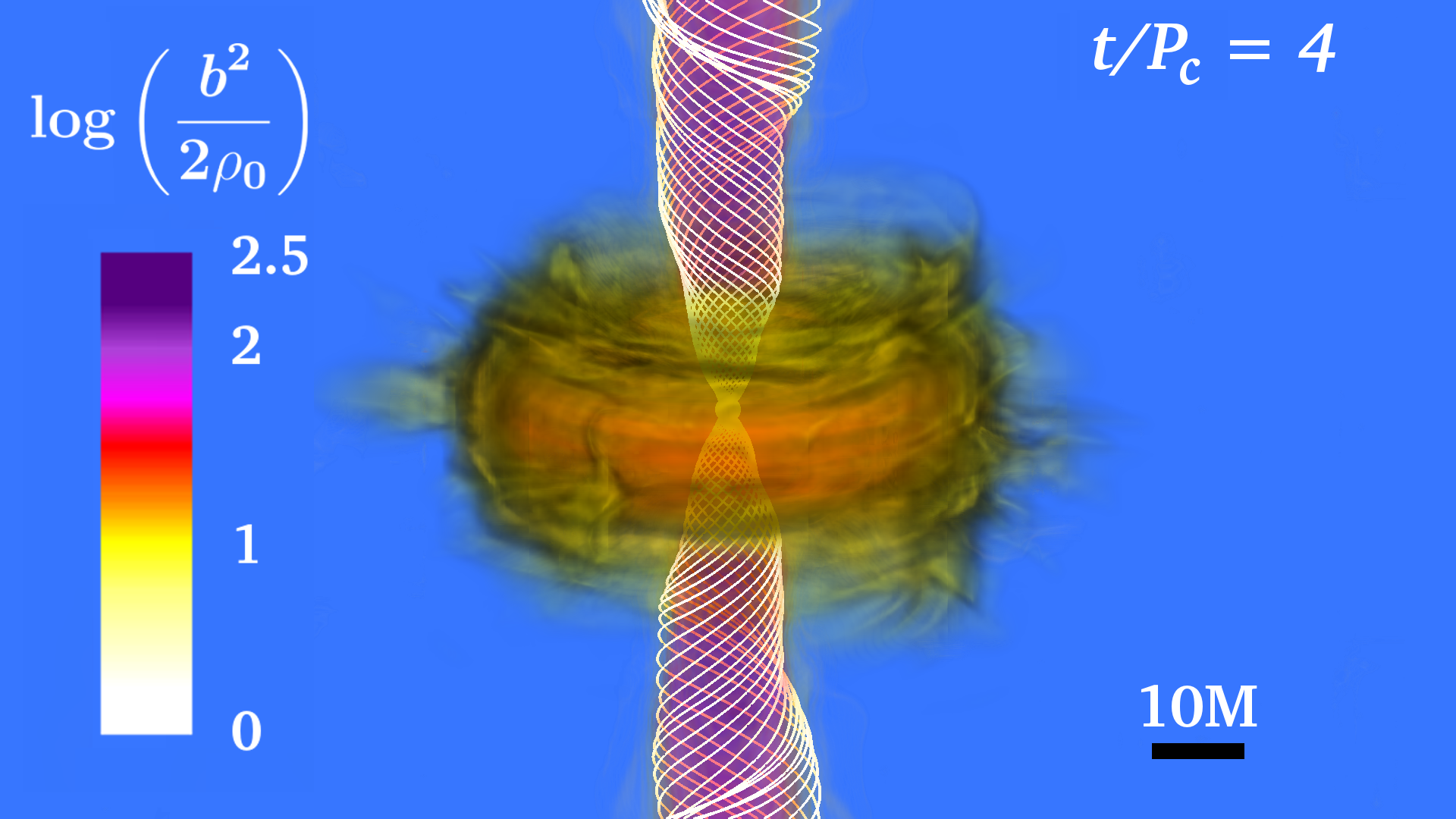}
\caption{Final rest-mass density profiles normalized to the initial maximum density (left column)
  and the force-free parameter inside the helical magnetic funnel (right column) for cases NS2 (top row), ES2 (middle
  row), and BH-disk (bottom row). White lines depict the magnetic field lines, while
  the arrows display fluid velocities, and $P_c$ is the rotation period measure at the point where
  the rest-mass density is maximum. Here $M = 5.9\ \rm km$.}
\label{fig:bfield}
\end{center}
\end{figure*}

To address the above questions, we employ our recently constructed, dynamically stable ergostars 
\cite{Tsokaros:2019mlz,Tsokaros:2020qju} and perform a
series of GRMHD simulations that include differentially rotating HMNSs  with and without ergoregions
to assess the effect of the ergoregion in launching a jet. At the same time, we make critical
comparisons between candidate jets and the ones generated by a BH-disk. Our results 
suggest that an ergoregion does not facilitate or inhibit the launching of outflows. A magnetically-driven
outflow with a maximum value of $\Gamma_L\sim 2.5$ is launched whether or not an
ergoregion is present (see left column in~Fig.~\ref{fig:bfield}). This outflow therefore is not consistent
with sGRBs, which require flows to reach $\Gamma_L\gtrsim 20$~\cite{Zou2009}. In contrast to the BH-disk where the
force-free parameter~$B^2/8\pi\rho_0$ above the BH poles reaches values of $\gtrsim 100$, the
force-free parameter
in the HMNS cases is only~$\lesssim 10$. We find that the Poynting luminosity of our evolved
models is comparable to the values reported in \cite{Shibata_2011}, as well as to the BZ luminosity.
This shows that the value of the luminosity by itself cannot be taken as a criterion to distinguish
compact objects with masses in the range $3-5\ M_\odot$ (the so-called mass-gap).
Finally, the angular frequency ratio of the magnetic field lines around the HMNSs is at least
twice the value $\Omega_F/\Omega_H\sim 0.5$ expected from the BZ mechanism~\citep{Blandford1977}.
Thus our current simulations suggest that the BZ mechanism for launching relativistic jets
only operates when a spinning BH is present, and that neither normal NSs nor ergostars can be the 
central engines that power sGRBs.

\begin{center}                                                                  
  \begin{table*}
    \caption{Equilibrium models. ER denotes the existence or not of an ergoregion. 
      Parameter $\hat{A}=A/R_e$, where $R_e$, the equatorial radius, determines the
      degree of differential rotation, $R_p/R_e$ is the ratio of polar to equatorial
      radius, $M_0$ is the rest mass, $M$ is the ADM mass, $J$ is the ADM angular 
      momentum, $T/|W|$ is the ratio of kinetic to gravitational energy, $P_c$
      is the rotational period corresponding to the central angular velocity
      $\Omega_c$, $\Omega_c/\Omega_s$ is the ratio of the central to the surface angular velocity,
      and $t_{\rm dyn}\sim 1/\sqrt{\GR}$ the dynamical timescale.}                       
    \label{tab:idmodels}       
    \begin{tabular}{lcccccccccccc}
      \hline\hline
      Model & EOS   & ER      & $\hat{A}^{-1}$ & $R_p/R_e$ & $M_0\ [\Ms]$   & $M\ [\Ms]$  & $R_e\ [{\rm km}]$    
& $J/M^2$   & $T/|W|$  & $P_c/M$ & $\Omega_c/\Omega_s$  & $t_{\rm dyn}/M$\\
\hline
NS1           & ALF2cc &  \xmark  & $0.2$         & $0.4688$ & $6.973$         &
$5.587$       & $12.55$& $0.8929$ & $0.2423$ & $25.21$  & $1.359$  & $6.6$  \\ \hline
ES1           & ALF2cc &  \cmark  & $0.2$         & $0.4531$ & $7.130$         &
$5.709$       & $12.49$& $0.9035$ & $0.2501$ & $24.18$  & $1.378$  & $6.5$  \\ \hline
NS2           & SLycc2 &  \xmark  & $0.3$         & $0.4688$ & $4.839$         &
$3.944$       & $8.873$& $0.8666$ & $0.2329$ & $21.66$  & $1.707$  & $6.7$  \\ \hline   
ES2           & SLycc2 &  \cmark  & $0.3$         & $0.4531$ & $4.930$         &
$4.017$       & $8.753$& $0.8759$ & $0.2403$ & $20.58$  & $1.743$  & $6.6$  \\ 
\hline\hline
    \end{tabular}
  \end{table*}
\end{center}                                                           
%

%
\textit{Numerical setup.}\textemdash
The HMNS initial data are constructed using the GR rotating equilibrium code described in
\cite{1992ApJ...398..203C} using two equations of state (EOSs). The first one is ALF2cc which was employed in 
\cite{Tsokaros:2019mlz} to find the first dynamically stable ergostars. It is based on the
ALF2 EOS \cite{Alford2005} where the inner region with rest-mass density 
$\GR_0\geq\GR_{0s}=\GR_{0\rm nuc}=2.7\times 10^{14}\ {\rm g/cm^3}$ is replaced by  
$P=\GS(\GR-\GR_s) + P_s$, where $\GS$ is a dimensionless constant,
$\GR$ is the total mass-energy density, and $P_s$ the pressure at $\GR_s$. Here we assume $\GS=1$, i.e. a
causal core.
Since the causal core starts at a relatively low density, $\GR_{0\rm nuc}$, the models based on this
EOS have density profiles that resemble the ones found in quark stars which exhibit a finite surface
density. The crust of the NS that follows the ALF2cc EOS is about $\sim 6\%$ of the equatorial radius.
These models, denoted by NS1 and ES1 in Table \ref{tab:idmodels},
have been presented in Table I of \cite{Tsokaros:2019mlz} (with names iA0.2-rp0.47 and iA0.2-rp0.45, respectively). 
They both belong to the sequence of stars having the same central rest-mass density 
$\GR_0=4.52\times 10^{14}\ {\rm g/cm^3}$ and are 
differentially rotating NSs with a $j$-const rotation law, $j(\Omega)=A^2 (\Omega_c - \Omega)$, where $\Omega_c$ 
is the angular velocity at the center of the star, and $\hat{A}=5$. 
NS1 has no ergoregion, while the adjacent model ES1, which rotates slightly faster, does.

The second EOS is SLycc2 \cite{Tsokaros:2020qju} and is based on the SLy EOS \cite{Douchin01} with 
the interior region having rest-mass density $\GR_0\geq 2 \GR_{0\rm nuc}$ replaced by the same causal EOS
as given above.
Here the causal core is further from the surface of the star, so a 
smoother transition to the surface is accomplished. Differentially rotating models with this EOS have 
been fully explored in \cite{Tsokaros:2020qju}. Here we pick two models, NS2 and ES2, that again belong in the 
sequence of central rest-mass density $\GR_0=7.82\times 10^{14}\ {\rm g/cm^3}$ and have 
$\hat{A}=3.\bar{3}$. They are slightly more differentially rotating than the ALF2cc models. Model
NS2 has no ergoregion, while the adjacent model ES2 does. Further properties of all our adopted models
can be found in Table \ref{tab:idmodels}. In the following, we consider both pure hydrodynamic and magnetized
(MHD) evolutions of these models.

For the magnetized cases, the stars are initially threaded by a dipole-like magnetic field whose
strength at the NS poles ranges between $4.5\times 10^{14}\rm G$ to $1.5\times 10^{16}\rm G$,
and with a magnetic dipole moment aligned with the direction of the spin of the star
(see~top panel~of Fig.~3~in~\cite{Ruiz:2019ezy}). We verify that initially the magnetorotational-instability
(MRI) is  captured in our models by computing the  quality factor $Q_{\rm MRI}\equiv\lambda_{\rm MRI}/dx$,
which  measures the number of grid points per fastest growing (MRI) mode as in~\cite{UIUC_PAPER2}.
Here $\lambda_{\rm MRI}$ is  the fastest-growing MRI wavelength.
We choose astrophysically large magnetic fields to mimic their growth due to the Kelvin-Helmholtz
instability (KHI) and the MRI during BNS merger and HMNS
formation. These instabilities can amplify moderate magnetic fields ($\sim 10^{13} \rm G$) to rms  values in the HMNS
of~$\sim 10^{15.5}\rm G$, and locally up to $\sim 10^{17}\ \rm G$~\cite{Kiuchi:2015sga,Kiuchi:2014hja,Aguilera-Miret:2020dhz}. 
To capture the magnetosphere that surrounds the NS, we set a variable, low-density magnetosphere in the HMNS exterior
such that the plasma parameter $\beta\equiv P_{\rm gas}/ P_{\rm mag}=0.01$ everywhere~\cite{Ruiz:2018wah}.
In all of our cases, the low-density increases the total rest-mass of the star by $\lesssim 1\%$,
consistent
with the values reported previously (see~e.g.~\cite{Ruiz:2016rai}). The ideal GRMHD equations are
then integrated everywhere, imposing on top of the magnetosphere a density floor in regions where
$\rho_0^{\rm atm}\leq 10^{-10} \rho_0^{\rm max}$, where $\rho_0^{\rm max}$ 
is the initial maximum rest-mass density of the system.

In addition to the above HMNS models, we evolve a BH-disk of $4\,M_\odot$ with an initial dimensionless spin
$J_{\rm BH}/M^2_{\rm BH}= 0.9$ to match those parameters of models NS2 and ES2 in Table~\ref{tab:idmodels}. The 
BH is surrounded by a massless accretion disk modeled by a $\Gamma=4/3$ polytropic EOS which is
initially threaded by a pure poloidal magnetic field confined to the disk interior.  The maximum
value of $\beta$ is $10^{-3}$ (see~Eq.~2~in~\cite{Khan:2018ejm}). Since we neglect
the self-gravity of the disk, our model can be scaled to an arbitrary rest-mass
density and magnetic field, keeping $\beta$ constant (see Eq.~A4 in~\cite{Khan:2018ejm}). 

We evolve the above systems using the Illinois GRMHD moving-mesh-refinement code 
(see e.g.~\cite{Etienne:2010ui}), which employs the Baumgarte-Shapiro-Shibata-Nakamura 
formulation of the Einstein’s equations~\cite{shibnak95,BS} with  puncture 
gauge conditions~(see~Eq.~(2)-(4) in~\cite{Etienne:2007jg}). The MHD equations are solved in conservation-law form 
adopting high-resolution shock-capturing methods. Imposition of $\nabla\cdot\vec{B}=0$ 
during  evolution is achieved by integrating the magnetic induction equation using a 
vector potential  $\mathcal{A}^\mu$ \cite{Etienne:2010ui}). The generalized
Lorenz gauge \cite{Farris:2012ux} is employed to avoid the appearance
of spurious magnetic fields~\cite{Etienne:2011re}. Pressure is decomposed as a sum 
of a cold and a thermal part, $P = P_{\rm cold} + (\Gamma_{\rm th}-1)\GR_0 (\GE-\GE_{\rm cold})$
where $P_{\rm cold}, \GE_{\rm cold}$ are the pressure and specific internal energy 
as computed from the initial data EOS (ALF2cc or SLycc2). For the thermal part we 
assume  $\Gamma_{\rm th}=5/3$.

In our NS simulations we used nine nested refinement levels with minimum grid spacing (at the
finest refinement level) $\Delta x_{\rm min}=122\ \rm m$ for the ALF2cc models, while for the
SLycc2 models, whose radii are much smaller, we used a minimum resolution of $\Delta
x_{\rm min}=85.6\ \rm m$. In both cases the radius of the star is resolved by $\sim 102$ grid
points. For the BH-disk simulation we used eight refinement levels with 
$\Delta x_{\rm min}=0.034192 k^{3/2}\ \rm m$, where $k=P/\rho_0^\Gamma$ is the polytropic
constant in the initial cold disk.  The horizon radius is resolved by $\sim 41$ grid points.

%
%
\paragraph*{{\bf Pure hydrodynamic evolutions} \textemdash }
To probe the dynamical stability of models NS2 and ES2, we evolve them following the same procedure as
in~\cite{Tsokaros:2019mlz}, where the stability properties of models NS1 and ES1 were reported. We find
that these models remain in equilibrium for more than a hundred dynamical timescales ($\gtrsim 30$
rotation periods). 
Due to the large density close to the star's surface, centrifugal forces push the
outer layers of the star (low-density layers) slightly outwards, while the bulk of the star remains axisymmetric
to a high degree until the end of our simulations (see Fig.~1 in \cite{Tsokaros:2019mlz}).
We do not find evidence of any significant growing instabilities or outflows during this time.

%
%
\paragraph*{{\bf MHD evolutions} \textemdash }
Magnetically-driven  instabilities and winding inevitable change the differential rotation law in the
bulk of the stars, and ultimately lead to the transition of the HMNS into another state which may or
may not be dynamically stable, depending on the specific characteristics of the initial configuration and the
magnetic field.
In Fig. \ref{fig:bfield}, the magnetized evolution of the dynamically stable normal star NS2 and ergostar ES2
are shown in the top and middle rows respectively. In both cases after $30P_c$ the HMNSs are
still differentially rotating, although not in the same way as the initial configurations.
To probe the stability of magnetized ergostars and their EM characteristics
we survey HMNS models against different magnetic field configurations.
Notice that the 
Alfv\'en time for magnetic growth  in the HMNS 
(mainly by magnetic winding, followed by MRI) is 
$\tau_{\text{\tiny A}}\sim 10\,\rho_{14}^{1/2}\,B_{15}^{-1}\,R_{10}\rm ms$ 
(see~Eq.~2 in \cite{Ruiz:2020via}). Here $\rho_{14}=\rho/10^{14}\rm g/cm^3$ is the characteristic 
density of the HMNS, $B_{15}=B/10^{15}\rm G$, and $R_{10}=R/10\rm km$, where $R$ the stellar equatorial radius.

\begin{figure}
\includegraphics[width=0.99\columnwidth]{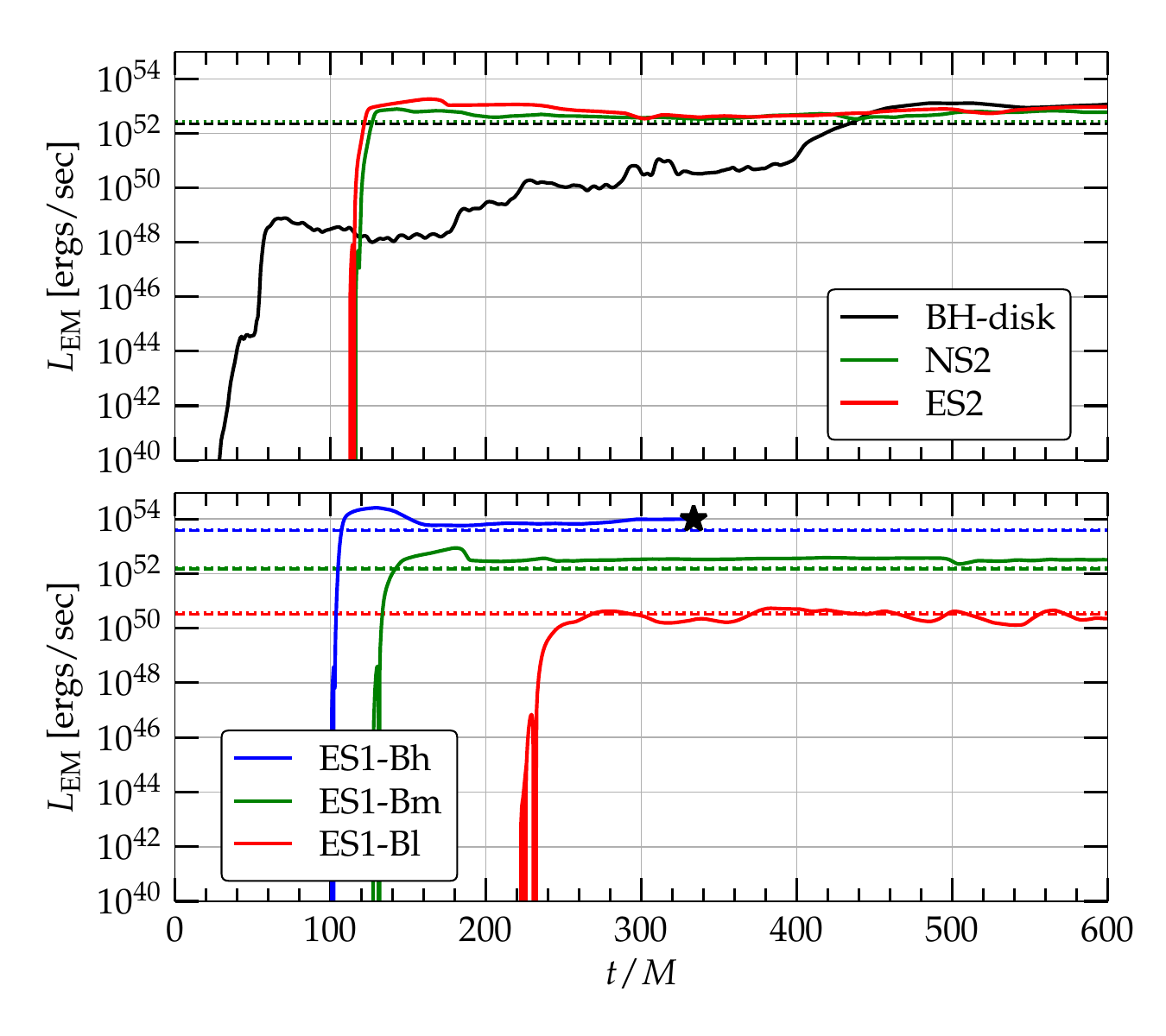}
\caption{Top panel: Outgoing EM Poynting luminosity for the three cases depicted in Fig.~\ref{fig:bfield}.
  Bottom panel: Outgoing EM Poynting luminosity for the ergostar ES1 and three different magnetic field 
  strengths. The black star marks the BH formation time.
  Dashed horizontal lines correspond to the BZ estimate is, while dotted horizontal lines (almost coincident
  with the dashed ones) to the estimate in Ref. \cite{Shibata_2011}.}
\label{fig:lum}
\end{figure}

The evolutions with the highest magnetic field strength, denoted by NS1-Bh and ES1-Bh, involve
the normal HMNS NS1 and ergostar ES1, which as discussed have similar physical properties 
(see Table~\ref{tab:idmodels}), and a poloidal magnetic field of  $1.5\times 10^{16}\rm G$.
We find that after $\sim \tau_{\text{\tiny A}}\sim 3P_c$  magnetic winding and MRI change the rotation
law of the HMNSs, driving the onset of stellar collapse. In the ES1-Bh case, the ergoregion expands
and after $4\tau_{\text{\tiny A}}\sim 12P_c$ an apparent horizon appears inside it.
A BH horizon in NS1-Bh forms at $\sim 5\tau_{\text{\tiny A}}\sim 15P_c$.
Using the isolated horizon formalism~\cite{dkss03}, we estimate that  the BH remnant in both cases
has a mass of $M_{\rm BH}\simeq 0.95M$ and dimesionless spin $a/M_{\rm BH}\sim 0.93$.
Here $M$ is the ADM mass of the corresponding HMNS (see Table~\ref{tab:idmodels}).
In both cases the BH is surrounded by an accretion disk with $\sim 4.2\%$ of the initial HMNS rest mass.
The Poynting luminosity ($L_{EM}\equiv -\int {T^r}_t\sqrt{-g}\,dS$) computed at different extraction
radii $r_{\rm ext}\gtrsim 80 M$) is $L_{EM}\sim 10^{54}\ \rm erg/s$ for both cases.
In the bottom panel of Fig.~\ref{fig:lum} the ES1-Bh is shown with a solid blue line and the black star
symbol marks the BH formation time, and the termination of our simulations. The fate of the BH-disk remnant
has  been already discussed in~\cite{Ruiz:2016rai}; as our initial  magnetic field is near
saturation, we do not expect additional enhancement  following BH formation. However, as
the matter above BH poles 
is accreted onto the BH, the ratio $B^2/8\,\pi \rho_0$  in the funnel will increase up to
values $\gtrsim 100$, and so the outflow can be accelerated to $\Gamma_L\gtrsim 100$ as required by
sGRB models~\cite{Zou2009}.

%
%

In the medium-magnetized case, ES1-Bm, where~$B=3\times 10^{15}\rm G$ ($\tau_{\text{\tiny A}}\sim 13P_c$), 
magnetic winding and the MRI
slowly drive the star into a new quasistationary configuration, which remains stable for more than $\gtrsim 30P_c$. 
We do not find evidence of any significant growing instabilities~\cite{Tsokaros:2019mlz}.
During this time, magnetic winding in the bulk of
the star induces a linear growth of the toroidal component of the magnetic field and corresponding  magnetic pressure.
Alfv\'en waves then propagate near the rotation axis transporting electromagnetic energy~\cite{Shibata_2011}.
This  builds up magnetic pressure above the stellar poles until eventually the inflow is halted and driven into
an outflow confined by the tightly wound field lines. The luminosity for this case is shown in the bottom
panel of Fig. \ref{fig:lum} with a solid green line. With a solid red line we also show the luminosity of
the lowest magnetized case ES1-Bl, whose initial magnetic field at the pole is $B= 4.5\times 10^{14}\rm G$. 

Similar to the evolution of ES1-Bm, when HMNSs NS2 and ergostar ES2 are threaded by a 
poloidal magnetic field of $5.25\times 10^{15}\rm G$ (Alf\'en time $\tau_{\text{\tiny A}}\sim 12P_c$)
they evolve stably for more than $30P_c$ (top and middle rows of~Fig.~\ref{fig:bfield}). 
In both cases, we observe that an outflow is launched after roughly $\sim 20P_c$ whether an ergoregion is 
present or not. As shown in Fig. \ref{fig:lum} top panel, the corresponding Poynting
luminosities are roughly the same, $L_{EM}\sim 10^{53}\rm erg/s$. 
{\it The comparison between NS2 and ES2 suggest that the ergoregion neither facilitates nor inhibits the launching 
of a magnetically-driven outflow}.

To assess if the BZ mechanism is operating in our systems, we compare the luminosity with that of a
BH-disk remnant that launches an incipient jet (right column in~Fig.~\ref{fig:bfield}). 
As seen in the top panel of Fig.~\ref{fig:lum}, the luminosity from the BH-disk matches the one coming from the stable HMNSs
NS2 and ES2. In the same panel we show with a dashed black line the luminosity predicted by the BZ mechanism: 
$L_{BZ}\sim 10^{51} (a/M_{\rm BH})^2(M_{\rm BH}/4M_\odot)^2\,B^2_{15}\ \rm erg/s$ \cite{BZeffect,MembraneParadigm}
for the BH-disk, which is consistent with the numerically computed one. 
We note here that if one naively applies the same formula to the HMNSs NS2 and ES2 one gets roughly
the same results, since the masses, dimensionless spins, and polar magnetic fields are the same as those of the 
BH-disk. The BZ estimates for the cases ES1-Bh, ES1-Bm, and ES1-Bl are shown with dashed lines in the bottom
panel of Fig.~\ref{fig:lum} which indicates agreement with the numerical values (solid lines).
In other words, the luminosity is not an efficient diagnostic for distinguishing outflows coming 
from a BH-disk or those coming from a NS. For example, a magnetized compact object in the mass gap will yield the
same luminosity as the BZ formula, making impossible its identification (BH-disk vs HMNS) through this criterion.

Ref. \cite{Shibata_2011} concludes that HMNSs like our models emit EM radiation 
with a luminosity $L_{EM}\sim 10^{51}B^2_{15}R^3_{10} \Omega_4\ \rm erg/s$, where $\Omega_4=\Omega/10^4\ \rm rad/s$. 
This estimate is shown in both panels of 
Fig.~\ref{fig:lum} with dotted lines, and coincide with the BZ dashed lines to a very high precision.
This is curious, since the two formula exhibit very different scaling with parameters.
Both of them are also close to the numerically computed luminosity.

One other possible source of luminosity is the pulsar spin-down luminosity~\citep{GJ1969}, which according
to \cite{Ruiz:2014zta} can be enhanced by as much as $\sim 35\%$ relative to its Minkowski value, if
general relativistic effects are taken into account. The numerical values shown in Fig. \ref{fig:lum} 
(as well the BZ estimate and the estimate from Ref. \cite{Shibata_2011}) show that our luminosities are at
least one order of magnitude larger than those found in~\cite{Ruiz:2014zta}. This is not surprising since:
(1) the stars in our recent analysis are differentially rotating, not uniformly rotating, as in \cite{Ruiz:2014zta}. 
Thus field lines tied to matter on the surface cannot ``corotate" with a rigidly rotating surface inside the light 
cylinder, as in a pulsar, but get wound up due to magnetic winding in the helical pattern we observe and which follows 
the exterior plasma flow. (2) Our magnetosphere is only marginally force-free, unlike the pulsars modeled in \cite{Ruiz:2014zta},
which are strongly force-free. Hence the exterior field topology,
which is always attached to the plasma, does not show nearly the same degree of winding under truly
force-free conditions and the flow is thus also with less winding.
Because of these reasons it is difficult to make a comparison regarding the spin-down luminosity in our simulations.
We will address this issue in a future work.

Following~\cite{prs15}, we measure the level of collimation of the outflow from the funnel opening angle, which
can be determined by the $B^2/8\pi\rho_0 \simeq 10^{-2}$ contour. Based on this value, we estimate an opening angle
of $\sim 25^\circ$ in our HMNS models. Such behavior has already been found in~\cite{Ruiz:2017due} where a HMNS
has been evolved for $\sim 200\ \rm ms$. Similar results have been reported in~\cite{Ciolfi:2020hgg}.
This level of collimation, although robust, is not as tight as in the
BH-disk where the opening angle is $\sim 15^\circ$. {\it The existence of the ergoregion does not seem to
affect either the development of this funnel structure or its geometry}.

In all cases, fluid elements inside the funnel have specific energy $E=-u_0-1 > 0$ and hence are unbound.
The characteristic maximum value of the Lorentz factor  is~$\Gamma_L \sim 2.5$
for cases NS2 and ES2, while~$\Gamma_L \sim 1.3$ in the BH-disk case. However, the force-free parameter 
$B^2/8\pi\rho_0$ inside the 
funnel in the latter case is a factor of $\gtrsim 10$ larger than that of the HMNS cases
(where $B^2/8\pi\rho_0\lesssim 10$). Since in steady-state the maximum attainable~$\Gamma_L$ of axisymmetric
jets equals the plasma parameter~\cite{B2_over_2RHO_yields_target_Lorentz_factor}, only in the BH-disk
case material inside the funnel can be accelerated to~$\Gamma_L\gtrsim 100$  as required for
sGRBs~\cite{Zou2009}.
Given the fact that magnetic winding and MRI will further transfer angular
momentum to the surface and make these HMNSs more uniformly rotating, leading 
to catastrophic collapse, as in case ES1-Bh, we do not expect that further evolution
will produce any significant changes \textit{until BH formation}, given what is shown in the right column
of Fig.~\ref{fig:bfield}. As the magnetic field is near saturation,
the only way for $B^2/8\pi\rho_0$ to grow is for the funnel to become baryon–free.
However, the outer layers of the star are a repository of matter that constantly supplies appreciable plasma 
inside the funnel.

Another characteristic of the BZ mechanism is the value~$\Omega_F/\Omega_H$. Here
$\Omega_F=F_{t\theta}/F_{\theta\phi}$ is the angular frequency of the magnetic field lines
and $\Omega_H=(a/m)(1+\sqrt{1-(a/m)^2})/(2m)$ is the angular frequency of the BH horizon.
For a Kerr BH with $a/m=0.9$ in a strongly force-free disk this ratio increases 
from $\sim 0.49$ at the pole to $\sim0.53$ at the equator \cite{Komissarov2001}. Numerical simulations that resulted 
in successful jet formation \citep{2004ApJ...611..977M,prs15,Ruiz:2016rai,Ruiz:2019ezy} have found this ratio to be 
$\sim 0.1-0.6$. 
We find $\Omega_F/\Omega_H\in [0.2,0.5]$ when $\GU\in [0,\pi/2]$ in our present BH-disk simulation.
Although this ratio is defined only for BHs, we nevertheless calculate its value using the central/surface
angular velocity of the neutron star.
We find $\Omega_F/\Omega_c \in [0.4,0.8]$, while 
if we normalize with the surface angular velocity $\Omega_s$ instead, the ratio $\Omega_F/\Omega_s$ becomes $\sim 1.5$ 
times larger. These results are approximately the same for both NS2 and ES2 and, therefore, the presence 
of the ergoregion does not seem to affect this ratio either.
Our preliminary conclusion is that for NSs this ratio can be at least twice as large as the one coming from the BHs.

%
\begin{figure}
\includegraphics[width=1.\columnwidth]{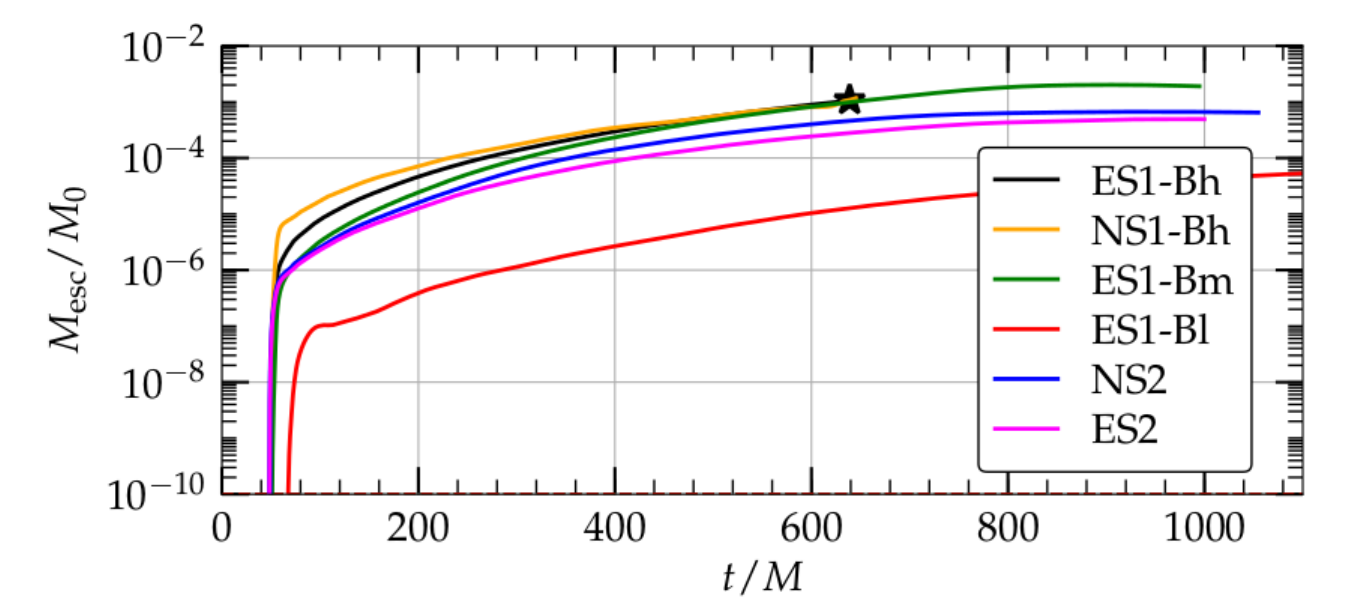}
\caption{Rest-mass fraction of escaping mass (ejecta) for the cases in Table~\ref{tab:idmodels} with
  different magnetic field strengths. The star marks the BH formation time for case ES1-Bh.}.
\label{fig:esc}
\end{figure}

For all cases we compute the ejected material 
$M_{\rm esc} = -\int\rho_0\GA u^t \sqrt{\GG}d^3x$ for $r>30M$ with specific energy $E>0$ and positive radial velocity.
We find that for cases NS1-Bh and ES1-Bh it is
$\sim 0.2\%$ of the total rest-mass of the HMNS~(see Fig.~\ref{fig:esc}), and therefore it may give
rise to an observable kilonova \cite{Metzger:2011bv}. Also, the ejecta coming from HMNSs NS1, and ES1
are approximately three times larger than those coming from HMNSs NS2 and ES2. The reason for that is that NS1, ES1
have a larger density close to the surface than  NS2 and ES2; therefore, when magnetic fields are present more
mass in the outer 
layers can get ejected.

\textit{Conclusions.}\textemdash
We surveyed different  HMNSs models with and without ergoregions with different initial strengths
of the seeded magnetic field to probe the impact of ergoregions on launching magnetically--driven outflows. 
We found that magnetized HMNSs launch a mildly relativistic jet confined inside a tightly-wound-magnetic-field
funnel whether or not an ergoregion is present. Our GRMHD simulations suggest that the properties of
the outflow, such as maximum Lorentz factors ($\Gamma_L\sim 2.5$), the plasma parameter ($B^2/8\pi\rho_0\lesssim 10$)
and the magnetic collimation, are not affected by the ergoregion. Notice that,
using force-free evolutions of magnetic fields on a \textit{fixed, homogeneous} ergostar background and based on the 
similarities between the topology of the strongly force-free EM fields and the induced currents on the ergostar vs. the ones on a
BH-disk, ~Ref.~\cite{Ruiz:2012te}  concluded that the BZ mechanism is likely to operate in ergostars.
As in~\cite{Ruiz:2012te} we do find some similarities in the topology of the magnetic fields (i.e.~collimation) 
for our ergostar and BH spacetimes, although we find the same similarities when a magnetized normal 
HMNS, instead of an ergostar, is compared. 
The Poynting luminosity in the HMNS  is comparable with that of the BH-disk remnant in which the BZ mechanism 
is operating, as well as with the luminosity reported by Ref. \cite{Shibata_2011}. Hence the luminosity
diagnostic cannot determine whether the BZ mechanism is operating or not.
On the other hand the ratio~$\Omega_F/\Omega_H$ turns out to be twice (at least) with the one computed in 
the BH-disk case. These results complement our previous studies with supramassive remnants \cite{Ruiz:2017due}
and suggest that in the hypermassive state it would be challenging for either normal stars or ergostars to be the
origin of relativistic jets and the progenitors of sGRBs.
Finally, we note here that similar maximum Lorentz factors and funnel 
magnetization have  been reported in simulations that include neutrinos \cite{Mosta:2020hlh}, so it is not clear if they
can play a significant role in the formation of jets.

Although threading an ergostar with a magnetic field is by itself inadequate to launch a bona fide BZ jet, there remain
mechanisms that can tap the rotational energy of the star by virtue of its ergoregion. These include the Penrose mechanism
\cite{Penrose:1969pc}. Noninteracting particles undergoing the Penrose process that can escape (e.g. various dark matter
candidates) carry off energy and angular momentum. Particles that interact with the NS matter and are captured conserve
(and redistribute) total angular momentum but can convert rotational kinetic energy into heat. Whether or not thermal
emission from this heat may be detectable, the lifetime of the ergoregion and any emission, and the fate of the ergostar
must all await further analysis.

Notice that the BZ mechanism represents the transfer of rotational kinetic energy from a Kerr BH to an outgoing Poynting
and matter flux.  Frame dragging above the poles tightly winds up the magnetic field  to force-free values, enabling it
to confine and drive accreting matter outward into a collimated jet. In a HMNS, the differentially rotating matter in
the star also serves to wind up and amplify the interior and exterior magnetic field, which becomes marginally force-free
outside the poles before saturating. This field also results in a Poynting flux and  some matter outflow from the surface,
again tapping the rotational kinetic  energy of the star. Our definition of BZ-like mechanism. The ergoregion, buried
inside the stellar surface, appears to play no significant  role in amplifying this mechanism. In a typical pulsar, by
contrast,  the NS is uniformly rotating, creating a  co-rotating, highly force-free magnetosphere insider the light cylinder, 
beyond which lines open up and contribute to an outgoing Poynting flux.  The stellar rotation and magnetic field induces
a strong electric field capable of stripping matter off the pulsar surface and  eventually populating the exterior with
tenuous plasma. Once again the  outflowing Poynting carries off rotational kinetic energy from the star. The rate of EM
dipole emission (and stellar spin-down) for a pulsar scales differently with stellar parameters than the rate
for EM emission in the BZ mechanism. We hope to explore the differences further in future simulations. 

\vspace{0.5cm}
We thank K. Parfrey and J. Schnittman for useful discussions.
This work was supported by NSF Grants No. PHY-1662211 and No. PHY-2006066, 
and NASA Grant No. 80NSSC17K0070 to the
University of Illinois at Urbana-Champaign. This work made use of the
Extreme Science and Engineering Discovery Environment (XSEDE), which is supported by National
Science Foundation Grant No. TG-MCA99S008. This research is also part of the Frontera computing project
at the Texas Advanced Computing Center. Frontera is made possible by National Science Foundation
award OAC-1818253. Resources supporting this work were also provided by the
NASA High-End Computing (HEC) Program through the NASA Advanced  Supercomputing  (NAS)
Division at Ames Research Center.   
%
\bibliographystyle{apsrev4-1}        
\bibliography{references}            
\end{document}